\documentclass{elsart}
 \usepackage{epsfig}

\usepackage{amssymb}
\begin{document}
\begin{frontmatter}


\title{Dynamical anomalies and the role of initial conditions in 
the HMF model}

\author
{Alessandro Pluchino},
\author
{Vito Latora},   
\author
{Andrea Rapisarda*}\corauth[cor1]{Corresponding author: andrea.rapisarda@ct.infn.it}

\address{Dipartimento di Fisica e Astronomia,  Universit\'a di Catania,\\
and INFN sezione di Catania, Via S. Sofia 64,  I-95123 Catania, Italy}

\begin{abstract}
We discuss the role of the initial conditions 
for the dynamical anomalies observed in the quasi-stationary 
states of the Hamiltonian Mean Field (HMF) model. 
\end{abstract}

\end{frontmatter}

\section{Introduction}
\label{intro}
The Hamiltonian Mean Field (HMF) model is a system of
N fully-coupled inertial spins which has been  intensively studied
in the last years 
\cite{ant1,lat1,lat2,lat4,lh1,lh2,celia1,plud,plu1,plu2,cabral,moya}. 
The model is particularly important for the paradigmatic  anomalous behavior exhibited 
by its            
out-of-equilibrium dynamics.  
Motivated by recent papers \cite{yama1,yama2} in which such anomalies 
were conjectured to exist only for very special initial conditions, 
in this paper, we 
 show that  anomalous dynamics, and in particular 
 fractal-like structures in phase space,
 power-law decay of correlations and   superdiffusion 
are obtained for a large class of initial conditions.  
Our results indicate that  these  anomalous behavior represents  more the rule rather 
than the exception. 
Connections with Tsallis thermostatistics \cite{tsa0,tsa1,tsa2,cho} 
 are briefly addressed.

\section{The model}
\label{The model}
The model describes a system of N planar classical spins 
${\stackrel{\vector(1,0){8}}{s_i}}=(cos\theta_i, sin\theta_i)$ 
with unitary mass and with an infinite-range interaction \cite{ant1}. 
The Hamiltonian, in the ferromagnetic case,  can be written as
\begin{equation}
\label{eq1}
        H= K+V
= \sum_{i=1}^N  {{p_i}^2 \over 2} +
  {1\over{2N}} \sum_{i,j=1}^N  [1-cos(\theta_i -\theta_j)]~~,
\end{equation}
\noindent
 where ${\theta_i} \in[-\pi,\pi],$ is the  angle of the  $ith$ spin and $p_i$ the
corresponding 
conjugate variable representing the rotational velocity. Since the
modulus of each spin is unitary, we can also view  the system as 
N interacting particles moving on the 
unit circle. The standard order parameter of the model is the
magnetization M, defined as 
$
\label{eq2} M = {1\over{N}} | \sum_{i=1}^N
\stackrel{\vector(1,0){8}}{s_i} |~~.
$ 
The equilibrium solution of the model 
exhibits a second-order phase transition from a 
low-energy condensed (ferromagnetic) phase with magnetization  $M\ne0$, 
to a high-energy (paramagnetic) one, where the spins are homogeneously
oriented on the unit circle and $M=0$. The {\em caloric curve},
i.e. the dependence of the energy density $U = E/N$ on the
temperature $T$, is given by
$U = {T \over 2} + {1\over 2} \left( 1 - M^2 \right)
~$\cite{ant1,lat1}.
The critical point is at energy density $U_c=\frac{3}{4}$,
which corresponds  to a critical temperature $T_c=\frac{1}{2}$.
\\
The dynamics of the HMF model shows several anomalies before complete
equilibration. More precisely,
if we adopt the so-called $M1$ initial
conditions, i.e. $\theta_i=0$  for all $i$ ($M(0)=1$) and
velocities uniformly distributed ({\it water bag}), the results of
the simulations, in a special region of energy values ($\frac{1}{2}
<U<U_c$), show a disagreement with the equilibrium prediction for a
transient regime whose lifetime depends on the system size N \cite{lat4,lh2}. 
In such a regime, the system remains trapped in metastable quasi-stationary 
states (QSS)   characterized by a temperature lower than the equilibrium one, a 
vanishing value of  magnetization and   
 Lyapunov exponents \cite{lh1}, 
anomalous diffusion and long-range correlations \cite{lat2,lat4,plud}, 
very slow and glassy-like   dynamics \cite{celia1,plu1,plu2}.

\section{Out-of-equilibrium dynamics vs initial conditions }

In this Section we show, by means of a series of numerical simulations,  
that the majority of the dynamical anomalies of the QSS regime are present 
not only for M1 initial conditions (ic), but also when the initial magnetization
$M(t=0)$ is in the range $(0,1]$.
We concentrate on the energy value $U=0.69$, where the anomalies 
are more evident. 
The case M(0)=0 has been studied in Refs. \cite{plud,yama1,yama2} 
and corresponds to a stable stationary state of the Vlasov equation \cite{yama2}. 
Such a state is spatially homogeneous from the beginning, thus 
the force acting on each spin is zero since $t=0$ and  correlations are 
almost absent. This case   represents a limiting situation as it will be shown 
in the following. 
To prepare the initial magnetization in the range $0 \le M \le 1$, we distribute 
uniformly the  orientation angle of  the spins  into a variable portion of the unitary circle.
In this way we fix the potential energy and  
we assign   the remaining part of the total energy as kinetic energy by using the 
usual water bag uniform distribution for the momenta.   
For the details
on the numerical integration used see refs. \cite{lat1,lat2,lh1}.
%
%
\begin{figure}
\label{fig1}
\begin{center}
\epsfig{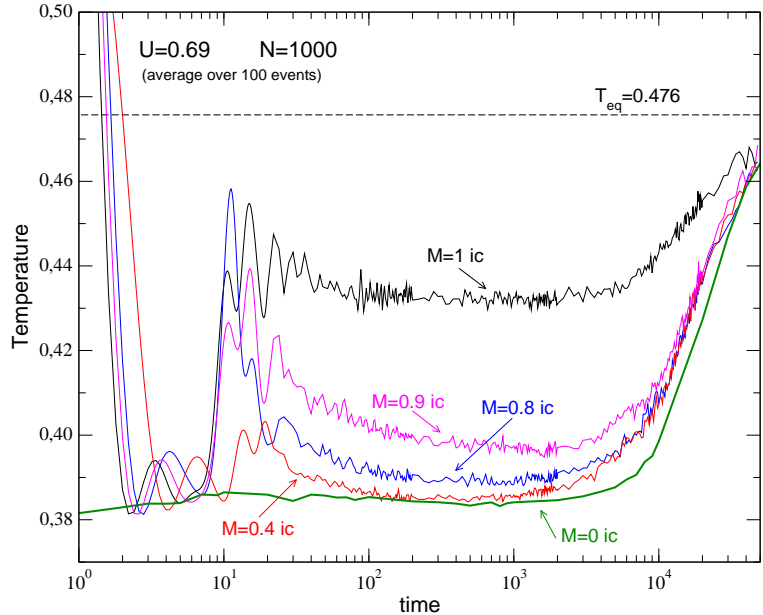}
\end{center}
\caption{Time evolution of the temperature for 
$U=0.69$, $N=1000$. Different initial conditions
of magnetization, ranging from $M=1$ to $M=0$, averaged over 100 events,
   are considered.
 }
\end{figure}
\begin{figure}
\label{fig2}
\begin{center}
\epsfig{figure=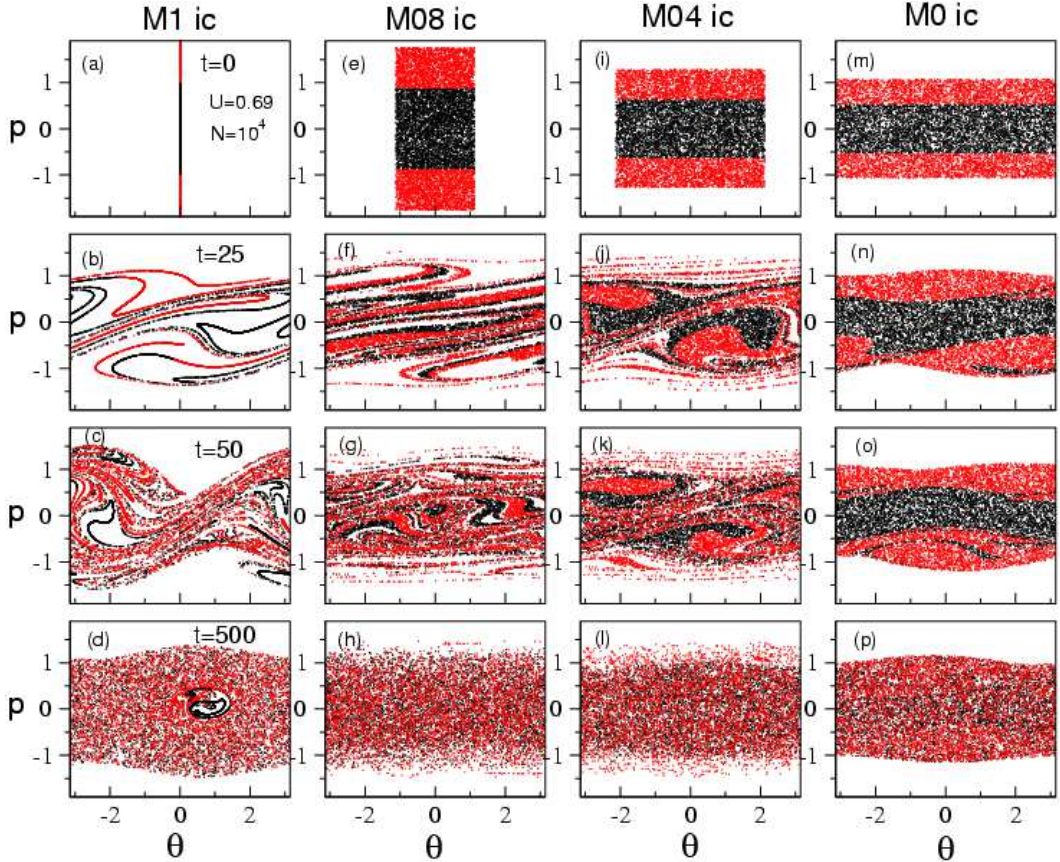,width=14truecm,angle=0}
\end{center}
\caption{
For  $U=0.69$, $N=10000$ we show a sequence of snapshots of the $\mu$-space 
at different times (from top to bottom). Four different initial magnetizations, 
namely $M=1,0.8,0.4,0$, are considered.
}
\end{figure}
%
%
\subsection{ Temperature plateaux and structures in phase-space}

In fig.1 we plot the time evolution of the temperature $T$, calculated by means of 
the average kinetic energy as $T=2<K>/N$. 
The simulations refer to $U=0.69 $, $N=1000$ and to 
different initial conditions. 
All the curves, except the one referring to  M=0  ic, show a fast relaxation from 
the  high 
initial temperature value. Then one observes small fluctuations around  
 a plateau region, before relaxation to equilibrium. 
We report as dashed line the equilibrium value.
The curves are obtained after an average over 100 events. Increasing 
this number fluctuations disappears completely.  The 
plateaux temperatures are  smaller than the equilibrium temperature and 
depend on the size and on the initial conditions used. 
No qualitative difference is found for greater sizes, apart from the 
fact that for $N \rightarrow \infty$ all the plateaux increase in time duration and 
tend to a limiting temperature $T_{N=\infty}=0.38$ (for $U=0.69$).
For any finite N the system always relaxes to the usual Boltzmann-Gibbs 
equilibrium value $T_{eq}=0.476$, although the relaxation time diverges 
linearly with N. Therefore the QSS regime can be considered a real equilibrium 
regime when the infinite size limit is taken before 
the infinite time limit \cite{lat4,lh2}.
\\
Although from the plot of T vs time the QSS behavior seems to be the same for all 
the initial conditions, this is not true for what concerns the correlations 
and their decay. 
%
%
\begin{figure}
\begin{center}
\label{fig3} \epsfig{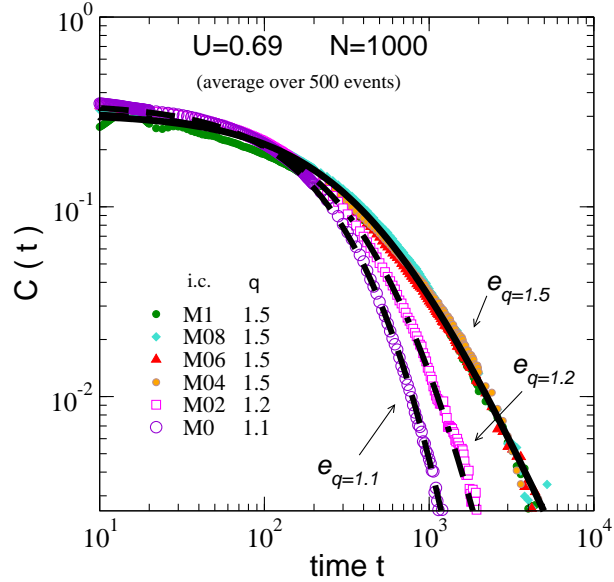}
\end{center}
\caption{Correlation functions vs time for different initial 
magnetizations (symbols). The curves are normalized q-exponentials 
defined by eq.(3).  
}
\end{figure}
%
In fig.2, for $U=0.69$ and  $N=10000$, 
we show the $\mu$-space of the system at different times and 
for four different initial magnetizations.  
The different colours distinguish  fast initial  spin velocities (in red) and slow 
ones (in black). 
While in the QSS regime the magnetization vanishes immediatly 
after the violent initial relaxation, 
dynamical structures, having fractal-like features \cite{lat4},  emerge in the $\mu$-space and then fade away only after a  long time.
These structures, that have been already studied in detail for 
M1 ic \cite{lat4,plud}, depend on the initial magnetization and thus on the initial value 
of the force, acting on each spin $j$, being $F_j=-M_x sin(\theta_j) + M_y cos(\theta_j)$. 
They are clearly visible for M08  and also for M04, while  
they are almost absent for M0 ic. 
  \subsection{Long-range correlations and anomalous diffusion}
In this section we focus on the decay of velocity correlations.
A quantitative way to estimate the velocity correlations is the
 autocorrelation function, defined as 
\begin{equation}
{C}(t)= {1\over{N}} \sum_{j=1}^N {p_j(t) p_j(0)} ~~.
\end{equation}
In fig.3 we plot the velocity autocorrelation functions for $N=1000$ and  
$M(0)={1, 0.8, 0.6, 0.4, 0.2, 0}$. Averages are taken over 500 different realizations.
The initial fast relaxation illustrated in fig.1 has not  been considered.
For $ M(0)\ge 0.4$ the correlation functions are very similar,
while the decay is faster for $M(0)=0.2$ and even more for $M(0)=0$.
The autocorrelation functions can be reproduced by means of the q-exponential
function 
\begin{equation}
e_q(z)= {\left[  1+(1-q) z \right]} ^{1\over(1-q)}
\end{equation}
proposed by Tsallis in his generalized thermodynamics \cite{tsa0,tsa1,tsa2}
with $z=-{x \over \tau}$. 
Here $\tau$ is a   characteristic time. Notice that one recovers the usual 
exponential decay for $q=1$ \cite{tsa0,tsa1,tsa2}.
In this way we can quantitatively discriminate  
between the  different initial conditions.
In fact we get a q-exponential with $q=1.5$ for 
$M\ge0.4$, while we get $q=1.2$ and  $q=1.1$ for $M=0.2$ and for $M=0$  
respectively. 
Thus for $M>0$ correlations exhibit  a long-range nature  and a slow decay: they 
are very similar, but  they 
diminish progressively below $M=0.4$ to become almost exponential for $M=0$.  
In ref. \cite{yama1} it was shown that this limiting case can been also fitted by a 
stretched exponential. 
%
\begin{figure}
\begin{center}
\label{fig3} \epsfig{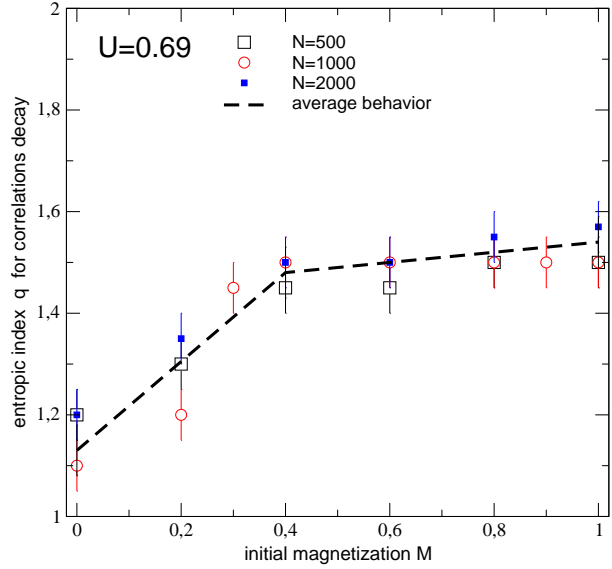}
\end{center}
\caption{Entropic index $q$ extracted from the decay 
of the correlation functions as a function of the initial magnetization  
and for several system sizes. 
The dashed line is the average behavior. 
}
\end{figure}
%
\begin{figure}
\begin{center}
\label{fig3} \epsfig{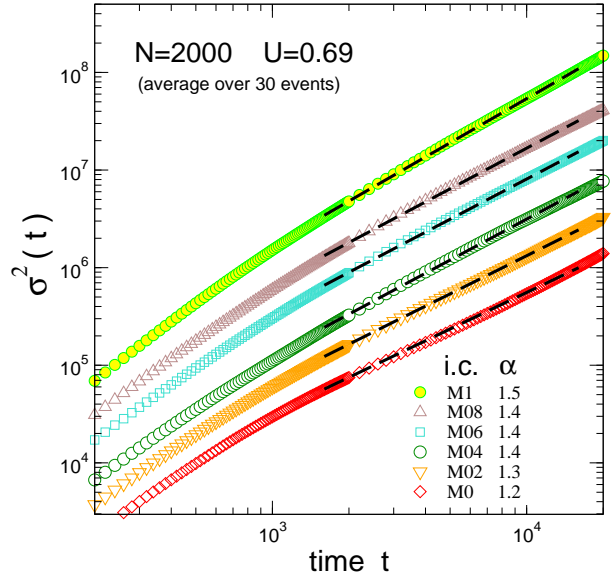}
\end{center}
\caption{We plot the mean square displacement of the angular motion $\sigma^2\propto 
t^{\alpha}$ \cite{lat2}
vs time for  different initial magnetizations. The exponent $\alpha$ which characterizes the 
behavior in the  QSS regime  and in its successive decay is also reported. The dashed
lines have  a slope  corresponding   to   these values.  
}
\end{figure}
%
%
In fig.4 we plot the q-values obtained for different initial conditions and 
different sizes of the system. It is possible to observe that the increase
of q from 1.1 to 1.5  with the initial magnetization
and the almost constant value  for $M(0)>0.4$. 
is not much  dependent  on the size 
of the system. Thus,  
in general, long-range correlations are obtained for a wide spectrum 
of initial conditions, while, again, the case 
M0  seems to be a very  special one. 
The latter  has been studied in detail in 
refs \cite{yama1,yama2} and it has been proven to be 
a stationary solution of the Vlasov equation that tends 
to attract the QSS. 
On the other hand,  Tsallis nonextensive thermostatistics scenario  seems to 
be a better candidate to explain the dynamical anomalies observed for finite
initial  magnetization \cite{lh2,tsa1,tsa2,cho}. A further indication in this 
direction  is provided by the  correlation between 
the value of the entropic index $q$ and  the exponent $\alpha$ 
of anomalous diffusion   \cite{bukman,beck}. The latter occurs if the mean square displacement
 in angle $\sigma^2 \propto t^\alpha$  has an exponent $\alpha \ne 1$.
Superdiffusion ($\alpha>1$) has been  observed in the 
HMF model  for  M1 initial 
conditions \cite{lat2,lh2}. In the present  investigation  we have  checked  that 
even decreasing the initial magnetization the system continues to show 
superdiffusion.
We illustrate this behavior in fig.5, where one sees, after a ballistic 
regime  $(\alpha=2)$ proper of the initial  fast relaxation, that in 
the QSS  plateau region and afterwards, 
the system shows superdiffusion. The exponent goes progressively 
from $\alpha=1.4-1.5$ for $0.4<M(0)<1$ to  $\alpha= 1.2$ for M0. In the latter case 
we have checked that increasing the size of the system  diffusion tends to be normal
($\alpha=1$ for N=10000). 
In correspondence, as previously shown, the entropic index $q$ characterizing 
the correlation decay, diminishes from $1.5$ to $1.1$, in good agreement with
 the relationship
$\alpha=2/(3-q)$  \cite{bukman}.
A detailed study of this behavior will be reported elsewhere \cite{anext}. 

\section{Conclusions}

The results discussed in  this paper show that  dynamical anomalies 
such as the emergence of fractal-like structures in  $\mu$-space, 
q-exponential velocity correlations and superdiffusion,
previously observed only for M(0)=1 initial conditions 
and pointing towards Tsallis thermodynamics scenario, 
are  always present when the  initial magnetization is greater than zero. 
Conversely, starting with M(0)=0, the dynamics  produces a very peculiar kind of QSS, different 
from all the other cases.
A possible explanantion of this different behavior could be 
  the sudden initial quenching 
characterizing, with different intensities, the dynamics for finite  initial  
magnetization. This fast relaxation is absent only for M(0)=0, which represents
 a limiting case.
In general our results show that 
anomalous behavior is  more ubiquitous than previously supposed.

\noindent
We thank Constantino Tsallis for useful discussions and suggestions.


\end{document}